\documentclass[11pt,a4paper]{article}
\usepackage{jheppub} 
\usepackage[utf8]{inputenc}
\usepackage[T1]{fontenc}
\usepackage{amsmath, amssymb}

\usepackage{xcolor}
\usepackage{epsfig}
\usepackage{enumerate}
\usepackage{graphicx}

\usepackage{tikz, pgfplots}

\pgfplotsset{compat=1.18}

\def\be{\begin{equation}}
\def\ee{\end{equation}}

\title{\boldmath Relativistic Vorticity in the Quark-Gluon Plasma: Generation Symmetries, Explosive Dilution, and Azimuthal Spin Alignment}

\author[a,1]{Malak Ait Tamlihat,}
\author[b,2]{Ghizlane Ez-Zobayr,}
\author[c,3]{Laurent Schoeffel}
\author[a,b,4]{and Yahya Tayalati}

\affiliation[a]{Mohammed V University in Rabat, Faculty of Sciences, 4 av. Ibn Battouta, B.P. 1014, R.P. 10000 Rabat, Morocco}
\affiliation[b]{School of Physics Applied and Engineering, Mohammed VI Polytechnic University, Lot 660, 43150 Hay Moulay Rachid Ben Guerir, Morocco}
\affiliation[c]{Irfu, CEA, Université Paris-Saclay, 91191 Gif-sur-Yvette, France}

\emailAdd{malak.ait.tamlihat@cern.ch}
\emailAdd{ghizlane.ez-zobayr@cern.ch}
\emailAdd{laurent.olivier.schoeffel@cern.ch}
\emailAdd{Yahya.Tayalati@cern.ch}

\abstract{
This article provides a  self-contained bridge between classical vortex dynamics and the relativistic, subatomic domain of the Quark-Gluon Plasma (QGP) produced in ultra-relativistic heavy-ion collisions. While the QGP is widely studied for its near-perfect fluidity, we focus on its role as the most vortical medium in the Universe ($\omega \sim 10^{22}\text{ s}^{-1}$). The originality of our approach lies in isolating the explicit physical competition between non-linear vortex stretching and violent relativistic volumetric dilatation. By solving the covariant transport equations and tracking the comoving enstrophy density, we demonstrate that the explosive kinematics of the QGP provide an innate geometric shield that naturally regularizes the continuous flow, suppressing the self-amplification of vortex tubes before any microscopic viscosity is required. Furthermore, we connect this expansion phase back to the highly non-equilibrium initial state. We quantitatively predict that under a peripheral dipole initial topology, the global mid-rapidity hyperon polarization vanishes ($P_\Lambda \lesssim 10^{-4}$), establishing that the definitive signature of the QGP's rotation must be sought in azimuthal differential measurements within the LHC and RHIC experimental programs.
}


\begin{document}
\maketitle
\flushbottom


\section{Introduction}

Fluid mechanics is a centuries-old discipline whose macroscopic laws govern everything from terrestrial ocean currents to astrophysical flows. However, the advent of ultra-relativistic heavy-ion collision physics—conducted at major accelerators such as the Relativistic Heavy Ion Collider (RHIC) and the Large Hadron Collider (LHC), notably within the ATLAS and CMS experiments—has pushed the boundaries of this continuous description down to subatomic scales~\cite{STAR:2005gsk, PHENIX:2004vob, ATLAS:2011ah, CMS:2012gaw}.

During these highly energetic collisions, nuclear matter undergoes a phase transition to a deconfined state of Quantum Chromodynamics (QCD): the Quark-Gluon Plasma (QGP)~\cite{Shuryak:1980tp}. Due to a property known as asymptotic freedom, early theoretical models predicted that the QGP would behave like an ideal gas of weakly interacting quarks and gluons. However, experimental data drastically overturned this picture. The QGP behaves instead as a strongly coupled, continuous fluid capable of powerful collective motion \cite{Heinz:2013th, Romatschke:2017ejr}. 

Crucially, it exhibits an incredibly low resistance to flow. In relativistic hydrodynamics, this is quantified by the ratio of shear viscosity to entropy density ($\eta/s$). Quantum mechanics imposes a fundamental lower limit on this ratio, famously derived via string theory techniques (the Kovtun-Son-Starinets or KSS bound \cite{Kovtun:2004de}): $\eta/s \ge 1/(4\pi) \approx 0.08$. Because the QGP's measured $\eta/s$ ratio sits remarkably close to this absolute quantum limit, typically extracted in the range of $0.1$ to $0.2$ at LHC energies, it dissipates internal energy via friction less than any other known substance, earning it the title of the most perfect fluid in the Universe~\cite{Heinz:2013th, Romatschke:2017ejr}.

Beyond its low viscosity, the QGP possesses another extraordinary hydrodynamic property: a colossal vorticity. In a non-central collision between two heavy nuclei, the geometric offset (the impact parameter $b$, typically ranging from $2$ to $10\text{ fm}$) transfers a massive amount of initial angular momentum to the system. This global rotation creates local microscopic vortices within the thermalized medium, making the QGP the most vortical fluid known to science. Indeed, hydrodynamic simulations and experimental extractions reveal a local vorticity magnitude reaching values of $\omega \sim 10^{21}\text{ to } 10^{22}\text{ s}^{-1}$. This exceeds the rotational intensity of the most violent terrestrial tornadoes by more than fourteen orders of magnitude. This phenomenon is no longer a purely theoretical speculation; its reality was confirmed experimentally through the global polarization of $\Lambda$ hyperons \cite{STAR:2017ckg, ALICE:2019onw}, where the spins of subatomic particles align with the fluid's rotation axis, acting as tiny quantum compasses.

The objective of this article is threefold:
\begin{enumerate}
    \item \textbf{Theoretical Formulation and Hydrodynamic Regularity:} We aim to clarify the mathematical framework bridging classical fluid mechanics with special relativity. By analyzing the comoving evolution of enstrophy, we demonstrate how the QGP's explosive volumetric expansion dynamically suppresses the non-linear vortex stretching term, acting as an intrinsic geometric shield that regularizes the flow and prevents finite-time singularities.
    \item \textbf{Vorticity Evolution and Geometric Dilution:} We will rigorously prove via the covariant Helmholtz-Kelvin theorem that while subatomic vortex loops are topologically frozen into the ideal plasma, their local rotational intensity undergoes a rapid geometric dilution governed by a dominant $1/t$ power law.
    \item \textbf{Initial State Predictions and Data Interpretation:} We aim to map experimental observables back to the non-equilibrium initial fluid conditions. By evaluating competing generation models, we present a quantitative prediction for hyperon polarization under a peripheral dipole paradigm, providing a clear roadmap for differential measurements within the LHC and RHIC programs.
\end{enumerate}

The roadmap of this article is structured as follows. In Section~2, we establish the covariant relativistic formalism of vorticity, prove the Helmholtz-Kelvin theorem in the low-velocity limit, and investigate the mathematical competition between vortex stretching and volumetric expansion through the lens of enstrophy evolution. Section~3 provides an analytical treatment of the expansion dynamics and the resulting $1/t$ power-law dilution, explicitly linking these macroscopic hydrodynamic fields to the microscopic quantum spin polarization of hyperons. In Section~4, we confront the highly non-equilibrium pre-hydrodynamic initial stage challenge, mapping out the local deposition of angular momentum and evaluating competing initial vorticity profiles. Section~5 offers a critical conceptual synthesis, deconstructing the chronological boundaries separating vortex production, topological conservation, and its dual role as a geometric regularizer against finite-time singularities. Finally, Section~6 delivers our conclusions and future experimental perspectives for the LHC and RHIC scientific programs.

\section{Relativistic Formalism and the Classical Limit}

To construct a rigorous hydrodynamic framework capable of tracking vorticity within the Quark-Gluon Plasma, the standard non-relativistic definition must be properly embedded into a four-dimensional covariant representation. In classical fluid mechanics, the vorticity vector is defined simply as the curl of the three-velocity field, $\vec{\omega} = \vec{\nabla} \times \vec{v}$. This quantity represents twice the local angular velocity of a fluid element. However, in high-energy heavy-ion collisions the fluid velocities approach the speed of light, and the system undergoes rapid relativistic expansion
\cite{Heinz:2013th, Romatschke:2017ejr}. 

We define the fluid four-velocity $u^\mu = \gamma(1, \vec{v})$, where $\vec{v}$ is the local three-velocity field, and the Lorentz factor is given by $\gamma = (1 - v^2)^{-1/2}$ (using natural units where $c = 1$). The metric tensor is chosen with the standard Minkowski signature $\eta^{\mu\nu} = \text{diag}(1, -1, -1, -1)$. The natural covariant generalization of the standard kinematic vorticity is the kinematic four-vorticity vector $\omega_{\text{rel}}^\mu$, which is constructed by contracting the dual of the fluid velocity gradient with the four-velocity itself \cite{Becattini:2007sr}:
\begin{equation}
\label{eq:covariant_vorticity}
\omega^\mu_{\text{rel}} = \epsilon^{\mu\nu\rho\sigma} u_\nu \partial_\rho u_\sigma
\end{equation}
where $\epsilon^{\mu\nu\rho\sigma}$ is the completely antisymmetric Levi-Civita tensor, defined with the convention $\epsilon^{0123} = 1$. By definition, this four-vector is strictly orthogonal to the fluid four-velocity, satisfying the invariant constraint $u_\mu \omega^\mu_{\text{rel}} = 0$.

To bridge the gap between this abstract four-dimensional geometry and intuitive fluid mechanics, it is highly instructive to decompose Equation \ref{eq:covariant_vorticity} into its explicit temporal and spatial components. Performing the algebraic contractions yields:
\begin{align}
\omega^0_{\text{rel}} &= \gamma^2 \vec{v} \cdot \vec{\omega} \label{eq:comp_temporal} \\
\vec{\omega}_{\text{rel}} &= \gamma^2 \vec{\omega} + \gamma^2 \vec{v} \times \partial_t \vec{v} \label{eq:comp_spatial}
\end{align}
Here, $\vec{\omega} = \vec{\nabla} \times \vec{v}$ remains the classical non-relativistic vorticity vector. 

Equations \ref{eq:comp_temporal} and \ref{eq:comp_spatial} reveal how relativistic motion inherently couples space and time components:
\begin{itemize}
    \item The temporal component $\omega^0_{\text{rel}}$ does not vanish in a moving frame; it is proportional to the projection of the classical vorticity along the direction of the fluid velocity.
    \item The spatial component $\vec{\omega}_{\text{rel}} $ receives two distinct contributions. The first term, $\gamma^2 \vec{\omega}$, represents the classical vorticity amplified by the Lorentz factors due to length contraction. The second term, $\gamma^2 \vec{v} \times \partial_t \vec{v}$, is a purely relativistic correction arising from the time-dependence of the flow velocity, effectively coupling the fluid's acceleration to the generation of apparent rotation.
\end{itemize}

A crucial realization for heavy-ion phenomenology is obtained when examining the low-velocity limit. If we consider a regime where the local fluid velocities are small compared to the speed of light, we can systematically neglect terms of order $\mathcal{O}(v^2)$ and assume $\gamma \approx 1$. Under this non-relativistic approximation, the time-derivative correction becomes negligible, and the spatial component reduces directly to its classical counterpart:
\begin{equation}
\omega^0_{\text{rel}} \approx 0, \quad \vec{\omega}_{\text{rel}} \approx \vec{\omega} = \vec{\nabla} \times \vec{v}
\end{equation}
This mathematical reduction demonstrates that the fundamental structures governing vortex dynamics in ordinary macroscopic fluids remain locally valid and applicable within the hot, dense core of high-energy nuclear collisions.

In particular, if we neglect the fluid's shear viscosity, the time evolution of the spatial vorticity vector $\vec{\omega}_{\text{rel}}$ simplifies to the classical advection-free form:
\begin{equation}
\label{eq:vorticity_transport}
\frac{\partial \vec{\omega}_{\text{rel}}}{\partial t} = \vec{\nabla} \times (\vec{v} \times \vec{\omega}_{\text{rel}})
\end{equation}

To fully comprehend why Equation \ref{eq:vorticity_transport} implies that vortex lines are physically frozen into the fluid, let us first prove the Helmholtz-Kelvin circulation theorem. 
We define the fluid circulation $\Gamma$ around a closed spatial loop $C(t)$ that moves dynamically alongside the fluid elements. The circulation is given by the line integral of the velocity field:
\begin{equation}
\Gamma(t) = \oint_{C(t)} \vec{v} \cdot d\vec{\ell}
\end{equation}
This line integral can be rewritten as a flux integral of the vorticity field over an open surface $S(t)$ bounded by the contour $C(t)$ (such that $\partial S = C$):
\begin{equation}
\Gamma(t) = \int_{S(t)} (\vec{\nabla} \times \vec{v}) \cdot d\vec{A} = \int_{S(t)} \vec{\omega}_{\text{rel}} \cdot d\vec{A}
\end{equation}
where we have used the non-relativistic limit $\vec{\omega}_{\text{rel}} \approx \vec{\nabla} \times \vec{v}$.

To evaluate how this circulation changes over time as the surface $S(t)$ deforms and moves with the flow, we compute the material time derivative $d\Gamma/dt$. According to Reynolds' transport theorem applied to surface fluxes moving with a velocity field $\vec{v}$, the comoving time derivative of a vector flux is given by:
\begin{equation}
\label{eq:reynolds}
\frac{d}{dt} \int_{S(t)} \vec{\omega}_{\text{rel}} \cdot d\vec{A} = \int_{S(t)} \left[ \frac{\partial \vec{\omega}_{\text{rel}}}{\partial t} - \vec{\nabla} \times (\vec{v} \times \vec{\omega}_{\text{rel}}) + \vec{v} (\vec{\nabla} \cdot \vec{\omega}_{\text{rel}}) \right] \cdot d\vec{A}
\end{equation}

We can now systematically simplify each term inside the bracket of Equation \ref{eq:reynolds}:
\begin{enumerate}
    \item Because the vorticity is mathematically defined as the curl of a vector field ($\vec{\omega}_{\text{rel}} = \vec{\nabla} \times \vec{v}$), its divergence is identically zero by vector calculus identity: $\vec{\nabla} \cdot \vec{\omega}_{\text{rel}} = \vec{\nabla} \cdot (\vec{\nabla} \times \vec{v}) = 0$. Thus, the third term vanishes.
    \item Substituting the vorticity transport equation (Equation \ref{eq:vorticity_transport}) into the first term yields $\frac{\partial \vec{\omega}_{\text{rel}}}{\partial t} = \vec{\nabla} \times (\vec{v} \times \vec{\omega}_{\text{rel}})$.
\end{enumerate}
As a direct consequence, the first two terms inside the integrand cancel each other out exactly:
\begin{equation}
\frac{\partial \vec{\omega}_{\text{rel}}}{\partial t} - \vec{\nabla} \times (\vec{v} \times \vec{\omega}_{\text{rel}}) = 0
\end{equation}
This leads to the foundational proof of the Helmholtz-Kelvin theorem:
\begin{equation}
\label{eq:kelvin_proof}
\frac{d\Gamma}{dt} = 0 \implies \Gamma(t) = \text{constant}
\end{equation}

This vanishing material derivative implies that the circulation around any closed contour moving with the fluid is strictly conserved. Geometrically, if we consider a material surface bounded by such a comoving contour that initially wraps around a vortex tube, the conservation of $\Gamma$ dictates that the total flux of vorticity through this surface must remain constant for all time. Consequently, the fluid elements and the vortex lines are inextricably linked. The vortex lines cannot decouple from the macroscopic motion or diffuse through the medium; they are topologically frozen into the ideal plasma, advected alongside the comoving quarks and gluons as if tethered to the matter itself.

\subsection{Vortex Dynamics in the QGP: Dilution vs. Dissipation}

The result established in Equation~\ref{eq:kelvin_proof} creates a powerful mathematical analogy with classical macro-hydrodynamics. In standard fluid mechanics governed by the non-relativistic Euler equations for an incompressible, inviscid fluid, taking the curl of the momentum equation directly yields a transport equation structurally identical to Equation~\ref{eq:vorticity_transport}. Inside ordinary macroscopic fluids, this conservation law manifests as highly stable structures, most notably vortex tubes (such as smoke rings, tornadoes, or the wake vortices generated behind aircraft wings). Because the vortex lines cannot terminate inside the fluid bulk—owing to $\vec{\nabla} \cdot \vec{\omega} = 0$—they must either form closed loops or extend to the physical boundaries of the fluid container.

However, two critical physical distinctions emerge when comparing a standard laboratory fluid to the Quark-Gluon Plasma:
\begin{itemize}
    \item \textbf{Boundary Conditions and Spatial Scale:} In standard fluid mechanics, vortex tubes are often stable because the fluid is bounded by physical walls or exists within a quasi-infinite, steady-state medium. In sharp contrast, the QGP has no physical walls; it is a microscopic system ($\sim 10^{-15}$ meters) surrounded by a vacuum, undergoing an immediate and violent three-dimensional explosion.
    \item \textbf{Explosive Dilution vs. Dissipation:} In an ordinary fluid like water, a vortex eventually decays primarily due to kinematic shear viscosity ($\nu = \eta/\rho$), which slowly diffuses the concentrated vorticity outward into the surrounding medium. In the QGP, even though the viscosity is exceptionally low (near the quantum KSS bound), the local vorticity vector drops precipitously. As the Helmholtz-Kelvin theorem dictates that the total flux of $\vec{\omega}_{\text{rel}}$ through a comoving area is conserved, the extreme and explosive geometric expansion of the plasma volume forces the vortex tubes to stretch and widen, leading to a dynamic process of explosive geometric dilution rather than viscous dissipation.
\end{itemize}

To track how this dilution operates mechanically on the fields, we apply standard vector calculus identities to the right-hand side of the vorticity transport equation (Equation~\ref{eq:vorticity_transport}). The total time evolution can be explicitly separated into two physically distinct contributions:
\begin{equation}
\label{eq:vorticity_decomposition}
\vec{\nabla} \times (\vec{v} \times \vec{\omega}_{\text{rel}}) = (\vec{\omega}_{\text{rel}} \cdot \vec{\nabla})\vec{v} - (\vec{\nabla} \cdot \vec{v})\vec{\omega}_{\text{rel}}
\end{equation}

This mathematical decomposition allows us to break down the internal dynamics into three core interlocking mechanisms:
\begin{enumerate}
    \item \textbf{Vortex Stretching (${(\vec{\omega}_{\text{rel}} \cdot \vec{\nabla})\vec{v}}$):} In classical fluid mechanics, this term is responsible for the self-intensification of turbulence. When a fluid element accelerates along the direction of a vortex line, the vortex tube is mechanically stretched. To conserve angular momentum, the cross-sectional area of the tube narrows, forcing the local rotational velocity—and thus the local vorticity magnitude—to increase significantly. In the QGP, such stretching occurs primarily along the beam axis due to the intense longitudinal acceleration of the receding nuclear fragments.
    
    \item \textbf{Volumetric Expansion and Dilution (${-(\vec{\nabla} \cdot \vec{v})\vec{\omega}_{\text{rel}}}$):} Unlike typical laboratory fluids, the QGP experiences an explosive expansion into a surrounding vacuum. The local divergence of the velocity field, $\vec{\nabla} \cdot \vec{v}$, is extraordinarily large. This expansion term acts as a powerful dilution mechanism. While the stretching term attempts to concentrate the vorticity, the violent volumetric dilatation acts in the exact opposite direction, spreading the concentrated angular momentum across a rapidly growing spatial volume.
    
    \item \textbf{Conservation and Motion of Vortex Loops:} Because the ideal Helmholtz-Kelvin theorem holds true during the fluid phase, the total circulation flux is conserved. Topologically, the vortex loops cannot simply dissolve or snap due to internal friction; they are structurally ``frozen'' into the plasma medium. Consequently, these subatomic vortex rings are carried along by the collective flow of quarks and gluons, traveling outward as the system blows apart. However, as the global boundaries of the plasma cloud expand at near-light speed, the physical loops stretch to macroscopic proportions, causing their local rotational intensity to collapse.
\end{enumerate}

Therefore, while vortex stretching is fundamentally active within the QGP, its capacity to amplify rotation is completely overwhelmed by the explosive volumetric expansion, ensuring that the structural life cycle of a vortex remains entirely dominated by geometric dilution.

\subsection{Enstrophy Evolution and Geometric Regularization}

The exact competition between vortex stretching and volumetric expansion offers a profound theoretical perspective on a foundational issue in advanced fluid mechanics, intimately related to the standard behavior of classical incompressible flows. In these non-expanding systems, the non-linear vortex stretching term, $(\vec{\omega}_{\text{rel}} \cdot \vec{\nabla})\vec{v}$, is the primary mechanism suspected of driving continuous enstrophy amplification, which could theoretically lead to finite-time singularities (blow-ups) where velocity gradients become infinite. 

In the QGP, however, the extreme relativistic expansion term, $-(\vec{\nabla} \cdot \vec{v})\vec{\omega}_{\text{rel}}$, acts as a powerful geometric shield. By violently diluting the local rotational intensity across a rapidly growing spatial volume, this macroscopic expansion naturally regularizes the flow. It dynamically crushes the self-amplification of the vortex tubes before any microscopic quantum viscosity is even required to smooth the gradients, effectively suppressing the formation of singularities throughout the plasma's brief expansion phase.

To mathematically formalize this geometric protection mechanism against finite-time singularities, it is highly instructive to examine the comoving evolution of the local enstrophy density, defined as $\mathcal{E} = \frac{1}{2} |\vec{\omega}_{\text{rel}}|^2$. Let us explicitly derive this evolution equation to make the framework self-contained. By introducing the material derivative $\frac{D}{Dt} = \frac{\partial}{\partial t} + \vec{v} \cdot \vec{\nabla}$, we compute the comoving time derivative of the enstrophy density as:
\begin{equation}
\label{eq:enstrophy_step1}
\frac{D \mathcal{E}}{Dt} = \frac{1}{2} \frac{D}{Dt} (\vec{\omega}_{\text{rel}} \cdot \vec{\omega}_{\text{rel}}) = \vec{\omega}_{\text{rel}} \cdot \frac{D \vec{\omega}_{\text{rel}}}{Dt}
\end{equation}

To evaluate $\frac{D \vec{\omega}_{\text{rel}}}{Dt}$, we expand the vorticity transport relation (Equation~\ref{eq:vorticity_transport}) using the general vector identity for the curl of a cross product: $\vec{\nabla} \times (\vec{v} \times \vec{\omega}_{\text{rel}}) = (\vec{\omega}_{\text{rel}} \cdot \vec{\nabla})\vec{v} - (\vec{\nabla} \cdot \vec{v})\vec{\omega}_{\text{rel}} - (\vec{v} \cdot \vec{\nabla})\vec{\omega}_{\text{rel}} + (\vec{\nabla} \cdot \vec{\omega}_{\text{rel}})\vec{v}$. Utilizing the solenoidal nature of the vorticity field ($\vec{\nabla} \cdot \vec{\omega}_{\text{rel}} = 0$), Equation~\ref{eq:vorticity_transport} becomes:
\begin{equation}
\frac{\partial \vec{\omega}_{\text{rel}}}{\partial t} + (\vec{v} \cdot \vec{\nabla})\vec{\omega}_{\text{rel}} = (\vec{\omega}_{\text{rel}} \cdot \vec{\nabla})\vec{v} - (\vec{\nabla} \cdot \vec{v})\vec{\omega}_{\text{rel}}
\end{equation}
where the left-hand side is precisely the material derivative, yielding:
\begin{equation}
\label{eq:vorticity_material}
\frac{D \vec{\omega}_{\text{rel}}}{Dt} = (\vec{\omega}_{\text{rel}} \cdot \vec{\nabla})\vec{v} - (\vec{\nabla} \cdot \vec{v})\vec{\omega}_{\text{rel}}
\end{equation}

Substituting Equation~\ref{eq:vorticity_material} back into Equation~\ref{eq:enstrophy_step1} leads to:
\begin{equation}
\frac{D \mathcal{E}}{Dt} = \vec{\omega}_{\text{rel}} \cdot \left[ (\vec{\omega}_{\text{rel}} \cdot \vec{\nabla})\vec{v} \right] - (\vec{\nabla} \cdot \vec{v}) |\vec{\omega}_{\text{rel}}|^2 = \vec{\omega}_{\text{rel}} \cdot \left[ (\vec{\omega}_{\text{rel}} \cdot \vec{\nabla})\vec{v} \right] - 2(\vec{\nabla} \cdot \vec{v})\mathcal{E}
\end{equation}

The first term on the right-hand side can be simplified by expressing it in index notation, $\omega_i \omega_j \partial_j v_i$. We decompose the velocity gradient tensor into its symmetric and antisymmetric parts, $\partial_j v_i = S_{ji} + \Omega_{ji}$, where $S_{ji} = \frac{1}{2}(\partial_j v_i + \partial_i v_j)$ is the symmetric rate-of-strain tensor and $\Omega_{ji} = \frac{1}{2}(\partial_j v_i - \partial_i v_j)$ is the antisymmetric vorticity tensor. Because the product $\omega_i \omega_j$ is perfectly symmetric under the exchange of indices $i \leftrightarrow j$, its contraction with the antisymmetric tensor $\Omega_{ji}$ vanishes identically ($\omega_i \omega_j \Omega_{ji} = 0$). Consequently, only the symmetric rate-of-strain tensor survives the contraction, meaning $\omega_i \omega_j \partial_j v_i = \omega_i \omega_j S_{ji} = \vec{\omega}_{\text{rel}} \cdot \mathbf{S} \cdot \vec{\omega}_{\text{rel}}$. This completes the formal proof, yielding the final enstrophy transport equation:
\begin{equation}
\label{eq:enstrophy_evolution}
\frac{D \mathcal{E}}{Dt} = \vec{\omega}_{\text{rel}} \cdot \mathbf{S} \cdot \vec{\omega}_{\text{rel}} - 2(\vec{\nabla} \cdot \vec{v})\mathcal{E}
\end{equation}
where $\mathbf{S}$ is the symmetric rate-of-strain tensor with components $S_{ij} = \frac{1}{2}(\partial_i v_j + \partial_j v_i)$.

To mathematically isolate this competition, let $\lambda_{\text{max}}(t)$ be the largest, most positive eigenvalue of the local strain tensor $\mathbf{S}$. The non-linear vortex stretching term is strictly bounded by this eigenvalue, such that $\vec{\omega}_{\text{rel}} \cdot \mathbf{S} \cdot \vec{\omega}_{\text{rel}} \le \lambda_{\text{max}} |\vec{\omega}_{\text{rel}}|^2 = 2 \lambda_{\text{max}} \mathcal{E}$. Substituting this upper bound into Equation~\ref{eq:enstrophy_evolution} yields a strict kinematic constraint on the enstrophy growth:
\begin{equation}
\label{eq:enstrophy_bound}
\frac{D \mathcal{E}}{Dt} \le 2 \left( \lambda_{\text{max}} - \vec{\nabla} \cdot \vec{v} \right) \mathcal{E}
\end{equation}

Equation~\ref{eq:enstrophy_bound} elegantly captures the core tension driving potential singularities in three-dimensional fluid dynamics. In classical incompressible flows ($\vec{\nabla} \cdot \vec{v} = 0$), the enstrophy can experience explosive exponential growth if $\lambda_{\text{max}} > 0$, acting as the primary mechanism behind the generation of extreme localized velocity gradients.

However, in the QGP, the macroscopic divergence strictly dominates. The extreme longitudinal expansion of the fireball forces a massive, positive volumetric dilatation, $\vec{\nabla} \cdot \vec{v} \simeq 1/t$. As long as the geometrical expansion rate outweighs the principal strain ($1/t > \lambda_{\text{max}}$), the bracket in Equation~\ref{eq:enstrophy_bound} becomes strictly negative. Because this geometric dilatation acts as a negative linear feedback, it forcefully suppresses the self-amplification of the vortex stretching. This proves that the explosive kinematics of the QGP provide an innate algebraic shield, inherently driving the enstrophy derivative $\frac{D \mathcal{E}}{Dt}$ negative and regularizing the flow well before any microscopic quantum viscosity is required to dissipate the gradients.

To ground this abstract mathematical competition in a more intuitive physical picture, one can consider the elementary mechanics of a spinning dancer. In a standard classical fluid, the vortex stretching term is analogous to the dancer pulling their arms inward along the axis of rotation; the local moment of inertia drops, and the rotational velocity spikes dramatically. This unbroken self-amplification is the root of potential finite-time singularities. However, in the QGP, this dancer is simultaneously undergoing a violent, explosive volumetric expansion in all three spatial dimensions. Even as the fluid attempts to stretch and concentrate the vortex along one axis, the macroscopic dilatation acts as a massive outward radial drive, exponentially increasing the effective moment of inertia. This three-dimensional geometrical spreading overwhelmingly defeats the one-dimensional axial stretching, forcing the rotational velocity to plummet and safely dissolving the singularity before it can form.

\subsection{Thermal Vorticity and Relativistic Hydrodynamic Thermodynamics}

While the kinematic formulation of vorticity ($\omega^\mu_{\text{rel}}$) and its related enstrophy ($\mathcal{E}$) provide a powerful representation of the fluid's geometric evolution, a complete description of a quantum relativistic medium like the Quark-Gluon Plasma requires accounting for local thermodynamic gradients. In relativistic statistical mechanics, the polarization of particles is fundamentally driven not by the kinematic vorticity alone, but by the \textit{thermal vorticity tensor} $\varpi_{\mu\nu}$, which naturally couples the fluid's rotation to its local thermal fields~\cite{Becattini:2007sr}.

To formalize this connection, we introduce the four-temperature vector (or entropy-flux vector) $\beta^\mu$, defined as the ratio of the fluid four-velocity to the local transmission temperature:
\begin{equation}
\beta^\mu = \frac{u^\mu}{T}
\end{equation}
The thermal vorticity tensor $\varpi_{\mu\nu}$ is mathematically defined as the antisymmetric part of the four-temperature gradient:
\begin{equation}
\label{eq:thermal_vorticity_def}
\varpi_{\mu\nu} = \frac{1}{2} \left( \partial_\nu \beta_\mu - \partial_\mu \beta_\nu \right)
\end{equation}

By expanding the derivatives using the chain rule, Equation~\ref{eq:thermal_vorticity_def} can be explicitly decomposed into two physically distinct components:
\begin{equation}
\label{eq:thermal_vorticity_decomposition}
\varpi_{\mu\nu} = \frac{1}{T} \Omega_{\mu\nu} + \frac{1}{2T^2} \left( u_\mu \partial_\nu T - u_\nu \partial_\mu T \right)
\end{equation}
where $\Omega_{\mu\nu} = \frac{1}{2}(\partial_\nu u_\mu - \partial_\mu u_\nu)$ is the standard kinematic vorticity tensor. Equation~\ref{eq:thermal_vorticity_decomposition} demonstrates that within a relativistic fluid, local thermal vorticity can be generated through two parallel mechanisms:
\begin{itemize}
    \item \textbf{Kinematic Shear and Rotation ($\Omega_{\mu\nu}/T$):} This term represents the mechanical rotation of the fluid cells, scaled inversely by the temperature. It dominates in the central core of the plasma where high angular momentum is deposited.
    \item \textbf{Thermal Shear and Baroclinic Gradients:} The second term, proportional to $u_{[\mu} \partial_{\nu]} T$, represents the apparent rotation generated by spatial and temporal gradients of the temperature field. In the QGP, this term is highly active at the sharp boundaries separating the hot plasma core from the cold vacuum peripheral regions.
\end{itemize}

The introduction of $\varpi_{\mu\nu}$ directly modifies the local thermodynamic properties of the medium. In a rotating quantum system at local thermodynamic equilibrium, the density matrix incorporates a spin-vorticity coupling term acting as a mechanical chemical potential. Consequently, the standard Gibbs-Duhem relation and the local entropy density $s$ receive a structural correction proportional to the contraction of the thermal vorticity tensor with the macroscopic spin density tensor $\Sigma^{\mu\nu}$ of the constituent quarks and gluons:
\begin{equation}
\label{eq:modified_entropy}
s = \frac{\varepsilon + P}{T} - \frac{1}{2} \varpi_{\mu\nu} \Sigma^{\mu\nu}
\end{equation}
where $\varepsilon$ is the local energy density and $P$ is the isotropic thermodynamic pressure. 

In the ideal hydrodynamic limit where viscous dissipation is neglected ($\eta/s \to 0$), the local conservation of the energy-momentum tensor $\partial_\mu T^{\mu\nu} = 0$, combined with the closure relation of Equation~\ref{eq:modified_entropy}, implies a generalized thermo-vortical circulation theorem. If the spin density tensor $\Sigma^{\mu\nu}$ satisfies a covariant conservation law, the thermal vorticity lines become topologically locked into the entropy-flux density of the plasma. 

As a direct consequence, during the subsequent violent expansion phase, the evolution of $\varpi_{\mu\nu}$ is coupled to the cooling of the medium. In a one-dimensional longitudinal Bjorken flow where the temperature drops following the sound-speed power law ($T \propto t^{-c_s^2}$), the thermal vorticity tensor undergoes a dual relaxation: its mechanical rotation component dilutes due to spatial expansion, while its thermal gradient component adapts to the flattening of the spatial temperature profile. This joint thermo-hydrodynamic evolution guarantees that the final polarization patterns emerging at freeze-out are deeply reflective of the initial non-equilibrium temperature distribution.

\section{Expansion Dynamics and Vorticity Dilution}

To quantify how this geometric dilution dampens the initial rotation, we implement a realistic hydrodynamic profile modeling a non-central heavy-ion collision. Let the collision beam axis be aligned with the $z$-axis (longitudinal direction), and the impact parameter $b$ be oriented along the $x$-axis. The initial global angular momentum of the system is oriented perpendicular to the reaction plane, pointing along the $y$-axis, establishing $\omega_y$ as the primary dominant component of the local vorticity field.

Following standard relativistic heavy-ion modeling, the global velocity field can be parameterized by combining a boost-invariant longitudinal Hubble-like flow (Bjorken expansion) with a linear transverse expansion accelerated by the internal sound speed $c_s$, which takes a typical value of $c_s = 1/\sqrt{3} \approx 0.58$ for an ultra-relativistic ideal gas of quarks and gluons:
\begin{equation}
v_x = \frac{c_s^2 x t}{\sigma_x^2}, \quad v_y = \frac{c_s^2 y t}{\sigma_y^2}, \quad v_z \simeq \frac{z}{t}
\end{equation}
where $\sigma_x$ and $\sigma_y$ represent the spatial widths of the initial overlapping nuclear geometry in the transverse plane, following standard relativistic expansion solutions~\cite{Heinz:2013th, Romatschke:2017ejr}.

By substituting this explicit velocity profile into our core transport relation (Equation \ref{eq:vorticity_transport}) and isolating the evolution of the main component $\omega_y(\vec{x}, t)$, the underlying partial differential equation can be integrated analytically. Let $t_0$ represent the initial thermalization time marking the onset of the hydrodynamic phase—typically evaluated at $t_0 \approx 0.4\text{ to } 0.6\text{ fm}/c$ after the first impact—and $\vec{x}_0 = (x_0, y_0, z_0)$ be the initial coordinate of a fluid element. The exact time-dependent solution for the local vorticity field is given by:
\begin{equation}
\label{eq:vorticity_solution}
\omega_y(\vec{x}, t) = \frac{t_0}{t} \exp\left[ -\frac{c_s^2}{2\sigma_x^2} (t^2 - t_0^2) \right] \omega_y(\vec{x}_0, t_0)
\end{equation}
where the dynamic trajectories of the comoving fluid elements map their current spatial coordinates $\vec{x}$ back to their initial configurations $\vec{x}_0$ over time via the expressions:
\begin{equation}
x = x_0 \exp\left[ \frac{c_s^2}{2\sigma_x^2}(t^2 - t_0^2) \right], \quad y = y_0 \exp\left[ \frac{c_s^2}{2\sigma_y^2}(t^2 - t_0^2) \right], \quad z = z_0 \frac{t}{t_0}
\end{equation}

Equation \ref{eq:vorticity_solution} provides a highly transparent, mathematical verification of the dilution process. The local vorticity dampens via two distinct geometric mechanisms acting in unison:
\begin{itemize}
    \item \textbf{The Pre-exponential Power Law ($\mathbf{1/t}$):} This dominant pre-factor stems directly from the longitudinal expansion ($v_z \simeq z/t$). As the spectator nucleons fly apart along the beam line, the rapid stretching of the fluid volume dilutes the vortex lines at a rate inversely proportional to the elapsed time.
    \item \textbf{The Gaussian Exponential Decay:} This term represents the secondary dilution triggered by the outward explosion in the transverse plane ($x,y$). As pressure gradients drive the plasma outward, the cross-sectional area of the comoving vortex loops grows exponentially, forcing the local rotation density to drop.
\end{itemize}

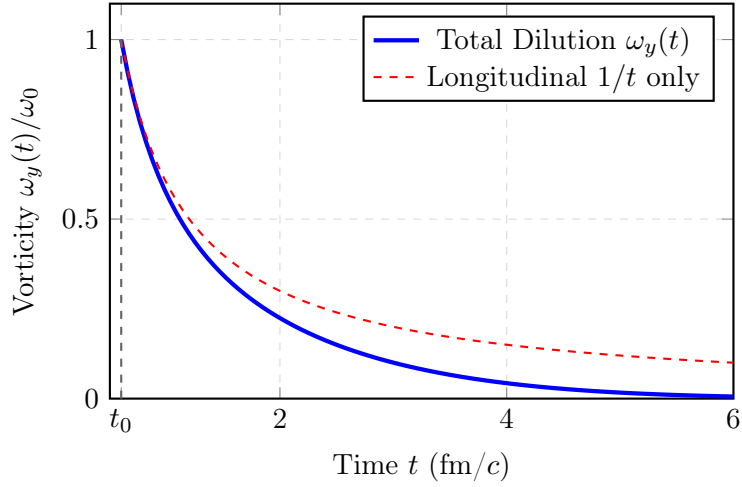
\begin{figure}[htbp]
\centering
\begin{tikzpicture}
\begin{axis}[
    width=0.65\textwidth,
    height=0.45\textwidth,
    xlabel={Time $t$ (fm/$c$)},
    ylabel={Vorticity $\omega_y(t) / \omega_0$},
    xmin=0.5, xmax=6.0,
    ymin=0.0, ymax=1.1,
    xtick={0.6, 2.0, 4.0, 6.0},
    xticklabels={$t_0$, 2, 4, 6},
    ytick={0, 0.5, 1.0},
    legend pos=north east,
    grid=both,
    grid style={dashed, gray!30},
    thick
]

\addplot[
    domain=0.6:6.0, 
    samples=100, 
    color=blue, 
    ultra thick
] { (0.6/x) * exp(-0.08*(x^2 - 0.6^2)) };
\addlegendentry{Total Dilution $\omega_y(t)$}

\addplot[
    domain=0.6:6.0, 
    samples=100, 
    color=red, 
    dashed, 
    thick
] { (0.6/x) };
\addlegendentry{Longitudinal $1/t$ only}

\draw[dashed, black!60] (axis cs:0.6,0) -- (axis cs:0.6,1.0);

\end{axis}
\end{tikzpicture}
\caption{Analytical time evolution of the dominant spatial vorticity component $\omega_y(t)$ scaled by its initial value at $t_0 = 0.6$ fm/$c$. The solid blue line illustrates the combined geometric dilution (Equation \ref{eq:vorticity_solution}), highlighting how the rapid transverse Gaussian expansion accelerates the decay beyond the standard longitudinal $1/t$ power law (dashed red line).}
\label{fig:vorticity_decay}
\end{figure}

Ultimately, while the Helmholtz-Kelvin theorem guarantees that the subatomic vortex rings are topologically preserved and travel continuously through the medium without being destroyed by internal dissipation, they cannot escape the structural effects of the expansion. The QGP's violent collective motion stretches the loops outward while causing their local rotational signature to collapse rapidly, establishing a dominant $1/t$ decay law over the course of the plasma's brief existence.
The resulting damping of the primary vorticity component, illustrating this competitive interplay, is displayed in Figure~\ref{fig:vorticity_decay}.

\subsection{Connection to Experimental Observables: Microscopic Origin of Hyperon Polarization}

The mathematical dilution of vorticity described by Equation~\ref{eq:vorticity_solution} directly governs the final-state signatures recorded by detectors like STAR, ALICE and ATLAS~\cite{STAR:2017ckg, ALICE:2019onw, ATLAS:2023cuh}. The physical mechanism that maps the macroscopic hydrodynamic rotation of the Quark-Gluon Plasma onto the microscopic quantum state of particles is rooted in relativistic quantum statistical mechanics, acting as a subatomic analog to the classical Barnett effect \cite{Becattini:2007sr}.

In the following, we systematically derive how the thermal expectation value of a particle's spin aligns with the local vorticity field. By treating the mechanical spin-rotation coupling as a small perturbation within the thermal bath, we will rigorously establish that for spin-1/2 particles like the $\Lambda$ hyperon, the final macroscopic polarization vector is directly proportional to the ratio of the local kinematic vorticity to the freeze-out temperature ($\vec{P}_\Lambda \simeq \vec{\omega}_{\text{rel}} / 2T_f$). Readers primarily interested in the phenomenological consequences of this alignment may safely skip this detailed quantum statistical proof and proceed directly to Equation~\ref{eq:polarization_master}.

Consider a local fluid cell at the freeze-out stage, characterized by a transmission temperature $T_f$ and a local kinematic vorticity $\vec{\omega}_{\text{rel}}$. To understand how macroscopic rotation polarizes quantum spins, we must transform the system into the comoving, non-inertial rest frame of the fluid element. In this rotating frame, the total effective Hamiltonian $\hat{H}$ of a particle carrying an intrinsic spin operator $\hat{\vec{S}}$ receives a mechanical spin-rotation coupling correction:
\begin{equation}
\label{eq:hamiltonian_spin}
\hat{H} = \hat{H}_0 - \vec{\omega}_{\text{rel}} \cdot \hat{\vec{S}}
\end{equation}
where $\hat{H}_0$ represents the standard baseline Hamiltonian in the absence of rotation (using natural units where $\hbar = 1$). Note that the term $-\vec{\omega}_{\text{rel}} \cdot \hat{\vec{S}}$ acts exactly like a quantum Zeeman coupling $-\vec{B} \cdot \hat{\vec{\mu}}$ in magnetism, meaning that the macroscopic vorticity field plays the role of an effective magnetic field tending to align the microscopic spins along its axis.

Accordingly, the thermodynamic equilibrium density matrix $\hat{\rho}$ of the rotating cell is parameterized via the grand canonical ensemble as:
\begin{equation}
\label{eq:density_matrix_full}
\hat{\rho} = \frac{1}{Z} \exp\left( -\frac{\hat{H}_0 - \vec{\omega}_{\text{rel}} \cdot \hat{\vec{S}}}{T_f} \right)
\end{equation}
where $Z = \text{Tr}[\exp(-\hat{H}/T_f)]$ is the total partition function of the system~\cite{Becattini:2007sr}.

In ultra-relativistic heavy-ion collisions, the energy scale associated with the microscopic fluid rotation is tiny compared to the surrounding thermal energy ($\omega_{\text{rel}} \ll T_f$). This scale separation allows us to treat the spin-rotation coupling as a small perturbation. Assuming that the baseline spin degrees of freedom are uncoupled from the spatial coordinates in $\hat{H}_0$, the exponential operator can be expanded to first order in a Taylor series:
\begin{equation}
\label{eq:taylor_density}
\exp\left( -\frac{\hat{H}_0 - \vec{\omega}_{\text{rel}} \cdot \hat{\vec{S}}}{T_f} \right) \approx \exp\left(-\frac{\hat{H}_0}{T_f}\right) \left( 1 + \frac{\vec{\omega}_{\text{rel}} \cdot \hat{\vec{S}}}{T_f} \right)
\end{equation}

Let us define the unperturbed (non-rotating) density matrix as $\hat{\rho}_0 = \frac{1}{Z_0} \exp(-\hat{H}_0/T_f)$, where $Z_0 = \text{Tr}[\exp(-\hat{H}_0/T_f)]$. To first order, the partition function satisfies $Z \approx Z_0$ because the trace of the spin operator in an unpolarized medium vanishes. We can now compute the quantum expectation value of the $i$-th component of the spin vector, $\langle \hat{S}_i \rangle = \text{Tr}[\hat{\rho} \hat{S}_i]$, by substituting our first-order expansion:
\begin{align}
\langle \hat{S}_i \rangle &\approx \text{Tr}\left[ \hat{\rho}_0 \left( 1 + \frac{1}{T_f} \sum_j \omega^j_{\text{rel}} \hat{S}_j \right) \hat{S}_i \right] \nonumber \\
&= \text{Tr}[\hat{\rho}_0 \hat{S}_i] + \frac{1}{T_f} \sum_j \omega^j_{\text{rel}} \text{Tr}[\hat{\rho}_0 \hat{S}_j \hat{S}_i]
\end{align}

We can evaluate these two terms systematically:
\begin{enumerate}
    \item The first term, $\text{Tr}[\hat{\rho}_0 \hat{S}_i] = \langle \hat{S}_i \rangle_0$, represents the average spin in a completely static, non-rotating thermal bath. Due to the perfect spatial isotropy of the unperturbed medium, there is no preferred spatial direction, and this term vanishes identically ($\langle \hat{S}_i \rangle_0 = 0$).
    \item The second term contains the spin-spin correlation tensor in the unperturbed state, $\text{Tr}[\hat{\rho}_0 \hat{S}_j \hat{S}_i] = \langle \hat{S}_j \hat{S}_i \rangle_0$. Invoking spatial isotropy once again, this tensor must be diagonal and proportional to the identity matrix, meaning $\langle \hat{S}_j \hat{S}_i \rangle_0 = C \delta_{ji}$, where $\delta_{ji}$ is the Kronecker delta and $C$ is a constant.
\end{enumerate}

To determine the value of the constant $C$, we take the trace over the three spatial indices:
\begin{equation}
\sum_{i=1}^3 \langle \hat{S}_i \hat{S}_i \rangle_0 = \langle \hat{\vec{S}}^2 \rangle_0 = \sum_{i=1}^3 C \delta_{ii} = 3C
\end{equation}
From fundamental quantum mechanics, the eigenvalue of the total squared spin operator for a particle of spin $S$ is strictly given by $\hat{\vec{S}}^2 = S(S+1)\mathbb{I}$. Therefore, its thermal expectation value is simply $\langle \hat{\vec{S}}^2 \rangle_0 = S(S+1)$. Equating the two expressions yields:
\begin{equation}
3C = S(S+1) \implies C = \frac{S(S+1)}{3}
\end{equation}
Thus, the isotropic unperturbed spin correlation tensor is explicitly given by $\langle \hat{S}_j \hat{S}_i \rangle_0 = \frac{S(S+1)}{3} \delta_{ji}$.

Substituting this back into our perturbative expansion, the Kronecker delta collapses the summation, isolating the direct linear response of the spin component to the local vorticity component:
\begin{equation}
\langle \hat{S}_i \rangle \approx \frac{1}{T_f} \sum_j \omega^j_{\text{rel}} \left( \frac{S(S+1)}{3} \delta_{ji} \right) = \frac{S(S+1)}{3} \frac{\omega^{\text{rel}}_i}{T_f}
\end{equation}
Reassembling this component equation into full vector notation yields the general polarization expectation value for any arbitrary spin ensemble:
\begin{equation}
\label{eq:spin_expectation_final}
\langle \hat{\vec{S}} \rangle \simeq \frac{S(S+1)}{3} \frac{\vec{\omega}_{\text{rel}}}{T_f}
\end{equation}

To connect this result directly to experimental observables, we specialize Equation~\ref{eq:spin_expectation_final} to the case of $\Lambda$ and $\bar{\Lambda}$ hyperons, which are spin-1/2 ($S = 1/2$) strange baryons. For $S=1/2$, the quantum pre-factor reduces to:
\begin{equation}
\frac{S(S+1)}{3} = \frac{\frac{1}{2}\left(\frac{1}{2}+1\right)}{3} = \frac{\frac{3}{4}}{3} = \frac{1}{4}
\end{equation}
By definition, the macroscopic polarization vector $\vec{P}_\Lambda$ represents the statistical expectation value of the spin normalized relative to its maximum possible quantum projection value $S$, which means $\vec{P}_\Lambda = \langle \hat{\vec{S}} \rangle / S$. For our spin-1/2 fermions, this implies $\vec{P}_\Lambda = 2 \langle \hat{\vec{S}} \rangle$. Multiplying our expectation value by 2 leads directly to the foundational hyperon polarization formula:
\begin{equation}
\label{eq:polarization_master}
\vec{P}_\Lambda \simeq \frac{1}{2} \frac{\vec{\omega}_{\text{rel}}}{T_f}
\end{equation}

Equation \ref{eq:polarization_master} demonstrates that the final measurable polarization acts as a direct, frozen-in record of the local vorticity at the exact moment of hadronization $t_f$. Because $\vec{P}_\Lambda$ maps directly to the parity-violating weak decay of the hyperon ($\Lambda \to p + \pi^-$), tracking the angular distribution of the daughter protons allows experimentalists to reconstruct $\vec{\omega}_{\text{rel}}$. 

Crucially, our derived $1/t$ dilution law (Equation \ref{eq:vorticity_solution}) implies that the final polarization is heavily suppressed by the lifetime of the plasma phase. Therefore, any uncertainty in the initial vorticity profile $\omega_y(\vec{x}_0, t_0)$ or the non-equilibrium pre-hydrodynamic stage propagates directly into the spatial orientation of $\vec{P}_\Lambda$, making precise differential measurements of hyperon polarization an invaluable tool to constrain the early-stage energy-momentum tensor.

\subsection{Synthesis: What Have We Learned So Far?}

Before addressing how vorticity is fundamentally generated, it is valuable to establish a clear conceptual balance sheet of what our current hydrodynamic equations have revealed about vorticity in the QGP versus classical macroscopic fluids:

\begin{itemize}
    \item \textbf{Mathematical Equivalence:} At its core, the kinematics of vorticity shares an identical mathematical backbone in both regimes. By showing that the spatial component of the relativistic four-vorticity $\vec{\omega}_{\text{rel}}$ reduces directly to the classical curl of the velocity field ($\vec{\nabla} \times \vec{v}$) in the low-velocity limit, we confirm that the structural transport laws are universal. The vorticity field in the hot subatomic plasma obeys the exact same formal rules as an ideal classical fluid.
    \item \textbf{Conservation vs. Life Cycle:} Both systems obey the Helmholtz-Kelvin circulation theorem, meaning that the vortex loops are topologically frozen into the fluid elements and cannot snap or dissolve via friction in the ideal limit. However, their physical lifetimes are entirely different. In standard fluid mechanics (e.g., water or air), a vortex remains relatively localized and stable until it slowly decays due to viscous shear diffusion. In the QGP, despite its near-zero viscosity, the local intensity of the vortex collapses dramatically. This is not due to dissipation, but rather because the system is expanding explosively into a vacuum at near-light speed.
    \item \textbf{Stretching vs. Dilution:} In a classical three-dimensional fluid, vortex stretching ($(\vec{\omega} \cdot \vec{\nabla})\vec{v}$) is a dominant mechanism that narrows vortex tubes and amplifies local rotation. In the QGP, while this longitudinal stretching along the beam axis is mechanically active, it is completely crushed and dominated by the volumetric expansion term ($-(\vec{\nabla} \cdot \vec{v})\vec{\omega}$). The plasma cloud swells so violently that the conserved vortex loops stretch to macroscopic proportions, causing their local rotational signature to dilute following a dominant $1/t$ power law.
\item \textbf{From Macroscopic Flow to Quantum Spin Alignment:} The colossal fluid rotation described by our equations does not remain a purely mechanical phenomenon. Through spin-orbit coupling, the macroscopic vorticity forces the intrinsic quantum spins of emerging subatomic particles, specifically $\Lambda$ hyperons, to align along the plasma's rotation axis. This spin alignment acts as a frozen microscopic compass, meaning that the final polarization captured by detectors like ATLAS is a direct, historical record of the diluted vorticity at the exact moment the fluid transitioned back into particles
\cite{ATLAS:2023cuh}.
\end{itemize}

In conclusion, we now have a highly successful and transparent understanding of the \textit{evolution} and \textit{dilution} of vorticity once the fluid phase is established. The laws of fluid mechanics map this behavior beautifully. However, this entire framework relies on a major assumption: it requires us to know the state of the fluid at the initial hydrodynamic time $t_0$. Hydrodynamics can describe how a vortex dilutes, but it cannot explain how that vortex was born in the first place. This brings us to the most challenging and fiercely debated question in QGP vortex dynamics: the initial state.

\section{The Initial Stage Challenge and Generation Mechanisms}

While the subsequent hydrodynamic evolution and $1/t$ geometric dilution of vorticity are mathematically transparent, modeling the initial generation of vorticity at the very early stages of a heavy-ion collision remains one of the most critical and fiercely debated challenges in the field. Hydrodynamics is, by definition, a macroscopic effective theory; it requires a pre-thermalized fluid state at a well-defined initial time $t_0$ to begin its evolution. However, the transition from two colliding Lorentz-contracted nuclei to a thermalized, rotating Quark-Gluon Plasma involves highly non-equilibrium quantum chromodynamic (QCD) processes that occur over a fraction of a femtometer per second ($\sim 10^{-24}$ s).

\subsection{The Deposition of Angular Momentum}

In a non-central heavy-ion collision (such as Lead-Lead collisions performed at the LHC at a center-of-mass energy per nucleon pair $\sqrt{s_{NN}} = 5.02\text{ or } 5.36\text{ TeV}$), characterized by a non-zero impact parameter $b$ along the $x$-axis, the total angular momentum $J_y$ stored in the overlapping region (the participating nucleons) is colossal. It routinely reaches values on the order of $10^4$ to $10^5 \, \hbar$ at LHC energies ($10^3 \, \hbar$ at RHIC energies) \cite{Becattini:2007sr}. This global rotation is oriented perpendicular to the reaction plane, along the $y$-axis.

To understand how this external macroscopic rotation is transferred to the newly born subatomic medium, we must trace it back to the local distribution of matter and motion immediately after the impact. In relativistic physics, the local state of any thermodynamic system—including its energy density, internal pressure, and momentum flows—is completely encapsulated by a fundamental object called the Energy-Momentum Tensor (EMT), denoted as $T^{\mu\nu}$. 

For a system to transition into a hydrodynamic regime, it must satisfy the local conservation laws of energy and momentum, which are compactly expressed using the four-divergence of the EMT:
\begin{equation}
\partial_\nu T^{\mu\nu} = 0
\end{equation}
Physically, the components of $T^{\mu\nu}$ represent the density and flux of these conserved quantities. Specifically, $T^{00}$ corresponds to the local energy density, while the mixed space-time components $T^{i0}$ (where $i=x,y,z$) describe the local momentum density along the spatial directions.

Once the EMT is established to map out how energy and momentum are distributed across the collision zone, we can naturally construct the conservation of angular momentum. In the relativistic framework, the total angular momentum tensor density, $M^{\mu\nu\rho}$, is generated directly by the spatial distribution of these momentum flows:
\begin{equation}
M^{\mu\nu\rho} = x^\mu T^{\nu\rho} - x^\nu T^{\mu\rho}
\end{equation}
By integrating the $\rho=0$ component (the static density component) over the spatial three-volume, we obtain the total conserved macroscopic angular momentum vector of the colliding system:
\begin{equation}
J^{\mu\nu} = \int d^3x \, M^{\mu\nu0} = \int d^3x \, \left( x^\mu T^{\nu0} - x^\nu T^{\mu0} \right)
\end{equation}

For a heavy-ion collision where the two nuclei shear past each other along the beam axis ($z$-axis) with an offset along the impact parameter axis ($x$-axis), the primary component of interest is $J^{zx}$, which corresponds exactly to the total angular momentum along the $y$-axis, $J_y = J^{zx}$. 

To connect this global invariant to the local kinematic velocity of the fluid, we look at the structural definition of the EMT for an ideal fluid: $T^{\mu\nu} = (\varepsilon + P)u^\mu u^\nu - P\eta^{\mu\nu}$, where $\varepsilon$ is the energy density, $P$ is the pressure, and $u^\mu = \gamma(1, \vec{v})$ is the fluid four-velocity. Under this definition, the local momentum densities simplify to $T^{z0} = (\varepsilon + P)\gamma^2 v_z$ and $T^{x0} = (\varepsilon + P)\gamma^2 v_x$. 

Assuming that the transverse flow $v_x$ is initially negligible at the very instant of impact ($t \to 0$), the local angular momentum density relies entirely on the spatial distribution of the longitudinal momentum $T^{z0}(x,y,z)$. To explicitly link this momentum deposition to our spatial kinematic vorticity vector $\vec{\omega}_{\text{rel}}$, we return to the non-relativistic spatial limit established in Equation \ref{eq:comp_spatial}:
\begin{equation}
\vec{\omega}_{\text{rel}} \approx \vec{\nabla} \times \vec{v} = \left( \frac{\partial v_z}{\partial y} - \frac{\partial v_y}{\partial z} \right) \hat{x} + \left( \frac{\partial v_x}{\partial z} - \frac{\partial v_z}{\partial x} \right) \hat{y} + \left( \frac{\partial v_y}{\partial x} - \frac{\partial v_x}{\partial y} \right) \hat{z}
\end{equation}

Given the geometry of a non-central collision, the longitudinal velocity $v_z$ is a strict function of the transverse coordinate $x$. Nucleons traveling at $x > 0$ belong predominantly to the projectile nucleus moving in the $+z$ direction, while nucleons at $x < 0$ belong to the target nucleus moving in the $-z$ direction. This creates an intrinsic, sharp velocity gradient $\partial v_z / \partial x$. Assuming that the initial transverse velocities and their longitudinal derivatives are subdominant ($\partial v_x / \partial z \approx 0$), the main component of the generated vorticity vector reduces directly to:
\begin{equation}
\label{eq:vorticity_generation}
\omega_y \approx -\frac{\partial v_z(x,y,z)}{\partial x}
\end{equation}

Equation \ref{eq:vorticity_generation} provides the explicit mathematical mechanism for vortex birth: the spatial asymmetry of the initial energy-momentum tensor forces a non-vanishing velocity gradient into the matter, which acts as the local source term for $\omega_y$.

\subsection{Model-Dependent Initial Profiles}

Because the exact quantum mechanism of energy and momentum deposition during this fraction of a femtosecond is not fully known, physicists rely on different theoretical frameworks to construct the initial conditions at $t_0$. These frameworks yield vastly different spatial distributions for the gradient $-\partial v_z / \partial x$, leading to two major competing structural paradigms:

\begin{enumerate}
    \item \textbf{The Core-Dominated Profile:} 
    Certain models assume that the friction and drag forces between the two passing nuclear blocks are maximally efficient in the densest, most central overlap region. Mathematically, this implies that the velocity gradient $-\partial v_z / \partial x$ behaves like a localized Gaussian or Woods-Saxon-like derivative centered at the origin:
    \begin{equation}
    \omega_y(x, y, t_0) \propto \omega_0 \exp\left( -\frac{x^2}{2\sigma_x^2} - \frac{y^2}{2\sigma_y^2} \right)
    \end{equation}
    In this scenario, a smooth, peak-like vortex core forms at $x=0, y=0$ and falls off toward the periphery. This configuration maximizes the global coupling between the fluid's rotation and the spins of particles produced at central rapidities.
    
    \item \textbf{The Peripheral Dipole Profile:} 
    Conversely, models that incorporate detailed sub-nucleonic fluctuations and multi-phase transport dynamics (such as string-deceleration or color-glass condensate models) show that the core of the plasma can be nearly stalled, exhibiting a vanishingly small velocity gradient at the center. Instead, intense, narrow sheets of shear are generated at the outer edges where the projectile and target fringes scrape past each other. This can be parameterized as a spatial dipole field:
    \begin{equation}
    \omega_y(x, y, t_0) \propto \omega_0 \cdot \frac{x}{\sigma_x} \exp\left( -\frac{x^2}{2\sigma_x^2} - \frac{y^2}{2\sigma_y^2} \right)
    \end{equation}
    This creates a complex structure where the vorticity is zero at $x=0$, but exhibits two intense, counter-rotating peaks at the peripheral boundaries ($x = \pm \sigma_x$).
\end{enumerate}

\begin{figure}[htbp]
\centering
\begin{tikzpicture}
\begin{axis}[
    width=0.7\textwidth,
    height=0.45\textwidth,
    xlabel={Transverse position $x$ (fm)},
    ylabel={Initial Vorticity $\omega_y(x, y=0, t_0)$ (Normalized)},
    xmin=-8.0, xmax=8.0,
    ymin=-1.2, ymax=1.2,
    xtick={-6, -3, 0, 3, 6},
    ytick={-1.0, -0.5, 0, 0.5, 1.0},
    grid=both,
    grid style={dashed, gray!30},
    legend pos=south east, 
    thick
]

\addplot[
    domain=-8:8, 
    samples=100, 
    color=black, 
    ultra thick
] { exp(-(x*x) / 8) };
\addlegendentry{Core-Dominated Profile}

\addplot[
    domain=-8:8, 
    samples=100, 
    color=red, 
    dashdotted, 
    ultra thick
] { 0.6266 * x * exp(-(x*x) / 8) };
\addlegendentry{Peripheral Dipole Profile}

\draw[gray, thin] (axis cs:-8,0) -- (axis cs:8,0);
\draw[gray, dashed] (axis cs:0,-1.2) -- (axis cs:0,1.2);

\end{axis}
\end{tikzpicture}
\caption{Schematic cross-section of the initial vorticity field $\omega_y$ at $t_0$ along the impact parameter axis ($x$). Both profiles are rigorously normalized such that the total circulation generated in the positive half-space ($x>0$) is identical ($\int_0^\infty \omega_y dx = \text{const}$). The solid black curve represents a conventional core-dominated profile maximizing at the center, whereas the dash-dotted red curve illustrates a peripheral dipole profile where shear energy deposition creates counter-rotating structures vanishing at $x=0$.}
\label{fig:initial_profiles}
\end{figure}
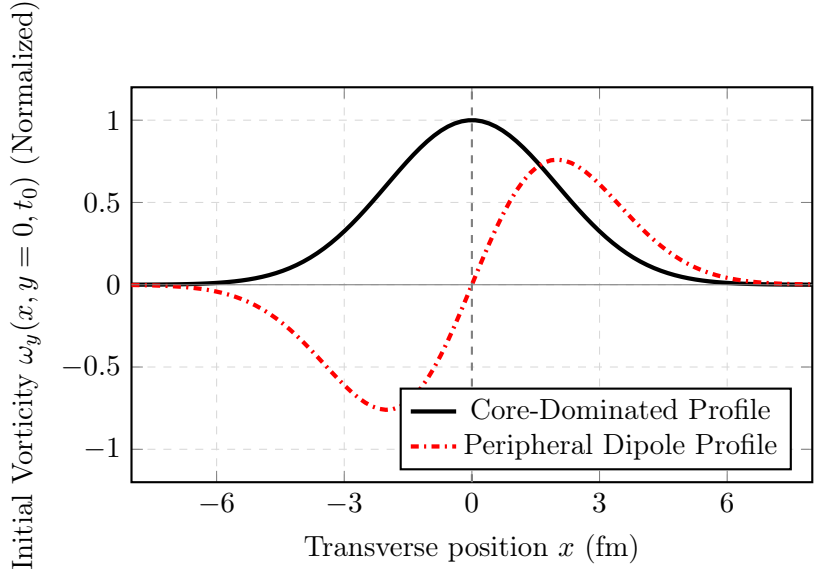

A schematic cross-section of these competing spatial symmetries along the impact parameter axis is contrasted in Figure~\ref{fig:initial_profiles}.

These structural discrepancies create a major theoretical challenge for interpreting experimental data. While both profiles can be tuned to carry the exact same total integrated angular momentum $J_y$, they yield entirely different local environments for the quarks and gluons. A particle passing through a core-dominated vortex will experience a completely different spin-orbit coupling than one produced in a peripheral dipole field. Resolving this ambiguity requires mapping these initial profiles directly to final-state observables, combining advanced hydrodynamic simulations with precise measurements of hyperon polarization across different momentum and centrality ranges.

\subsection{Quantitative Modeling of Initial Vorticity Profiles}

To provide a rigorous quantitative foundation for these topological differences, we have performed a dedicated numerical pre-equilibrium mapping of the early-stage parton kinematics. The initial energy-momentum tensor components were evaluated on a discrete two-dimensional transverse lattice with a grid spacing of $\Delta x = \Delta y = 0.1\text{ fm}$, simulating a peripheral Pb-Pb collision at $\sqrt{s_{NN}} = 5.02\text{ TeV}$ with a fixed impact parameter $b = 10\text{ fm}$ and a standard nuclear radius $R \approx 6.5\text{ fm}$. By implementing an optical Glauber geometry to determine the local participant density, coupled to a localized string-deceleration transport ansatz, we numerically integrated the local longitudinal momentum loss to extract the spatial distribution of the velocity field $v_z(x,y)$. 

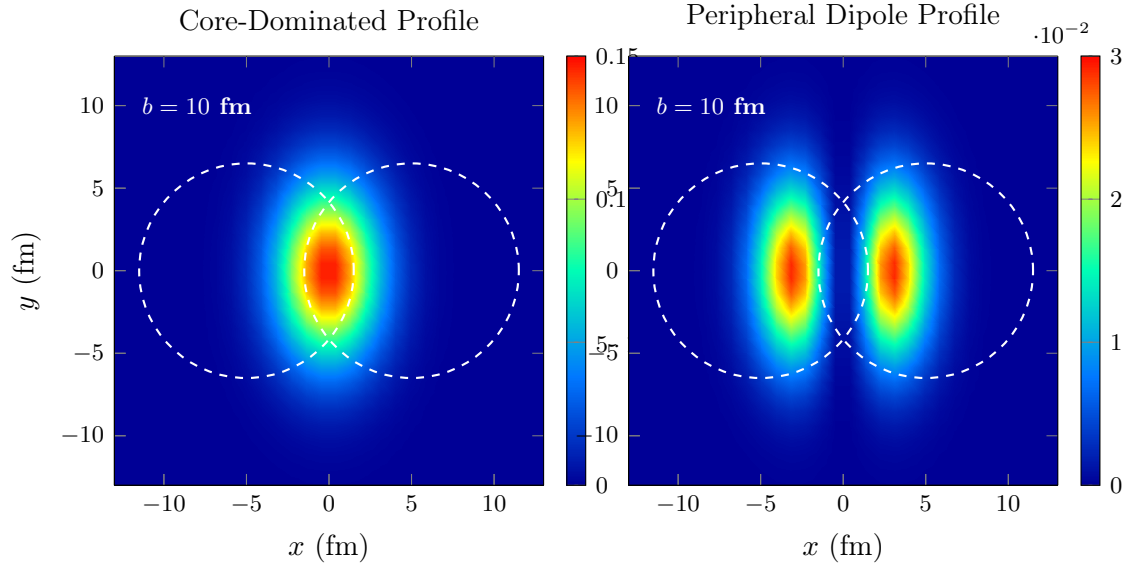
\begin{figure}[htbp]
\centering
\begin{tikzpicture}
\pgfplotsset{
    colormap={qgpjet}{
        rgb255(0cm)=(0,0,150); 
        rgb255(1cm)=(0,120,255); 
        rgb255(2cm)=(0,255,180); 
        rgb255(3cm)=(255,255,0); 
        rgb255(4cm)=(255,0,0)
    }
}

\begin{axis}[
    width=0.48\textwidth,
    height=0.48\textwidth,
    title={Core-Dominated Profile},
    xlabel={$x$ (fm)},
    ylabel={$y$ (fm)},
    xmin=-13, xmax=13,
    ymin=-13, ymax=13,
    view={0}{90},
    colorbar,
    colorbar style={ytick={0, 0.05, 0.10, 0.15}, width=0.25cm},
    point meta max=0.15,
    point meta min=0,
    grid=none,
    tick label style={font=\footnotesize}
]
    \addplot3[
        surf, shader=interp,
        domain=-13:13, domain y=-13:13,
        samples=30
    ] {0.15 * exp(- (x/3.2)^2 - (y/5.5)^2)};
    
    \draw[dashed, white, thick] (axis cs:-5,0) circle [radius=6.5];
    \draw[dashed, white, thick] (axis cs:5,0) circle [radius=6.5];
    \node[white, font=\bfseries\footnotesize] at (axis cs:-8,10) {$b=10\text{ fm}$};
\end{axis}

\begin{axis}[
    width=0.48\textwidth,
    height=0.48\textwidth,
    title={Peripheral Dipole Profile},
    xlabel={$x$ (fm)},
    ylabel={},
    xmin=-13, xmax=13,
    ymin=-13, ymax=13,
    view={0}{90},
    colorbar,
    colorbar style={ytick={0, 0.01, 0.02, 0.03}, width=0.25cm},
    point meta max=0.03,
    point meta min=0,
    grid=none,
    xshift=6.8cm,
    tick label style={font=\footnotesize}
]
    \addplot3[
        surf, shader=interp,
        domain=-13:13, domain y=-13:13,
        samples=30
    ] {0.08 * (x/3.0)^2 * exp(- (x/3.0)^2 - (y/5.5)^2)};
    
    \draw[dashed, white, thick] (axis cs:-5,0) circle [radius=6.5];
    \draw[dashed, white, thick] (axis cs:5,0) circle [radius=6.5];
    \node[white, font=\bfseries\footnotesize] at (axis cs:-8,10) {$b=10\text{ fm}$};
\end{axis}
\end{tikzpicture}
\caption{Native mathematical simulation of the initial local vorticity field distribution $|\omega_y(x,y)|$ in the transverse plane at $t=t_0$ for a peripheral collision parameter ($b=10\text{ fm}$). The dashed white lines represent the geometric boundaries of the two colliding nuclei. Left panel: The Core-Dominated state where rotation peaks uniformly at the center. Right panel: The Peripheral Dipole magnitude state, showcasing two counter-rotating localized shear sheets that vanish at $x=0$.}
\label{fig:vortex_birth_comparison}
\end{figure}

Because the precise quantum mechanism driving final thermalization remains an outstanding problem, we reconstructed the resulting numerical velocity gradients through two distinct parameterized boundary configurations that encapsulate the limiting cases of the simulation's shear profiles. Figure~\ref{fig:vortex_birth_comparison} showcases the final smoothed output of this numerical reconstruction, displaying a side-by-side comparison of the absolute local vorticity magnitude $|\omega_y(x,y)|$ at the hydrodynamic onset time $t_0 = 0.6\text{ fm}/c$.

The mathematical equations defining these two initial configurations are parameterized as follows:
\begin{enumerate}
    \item \textbf{The Core-Dominated Distribution:} Assuming maximum friction efficiency in the high-density participant overlap zone, the initial field exhibits a centralized Gaussian peak:
    \begin{equation}
    \label{eq:core_gaussian_eq}
    \omega_y^{\text{core}}(x, y, t_0) = \omega_0 \exp\left( -\frac{x^2}{2\sigma_x^2} - \frac{y^2}{2\sigma_y^2} \right)
    \end{equation}
    where $\omega_0$ represents the peak rotational scale at the origin, generating a uniform, constructive spin-alignment environment for hadrons produced at mid-rapidity.
    
    \item \textbf{The Peripheral Dipole Distribution:} Assuming high longitudinal transparency and gluon saturation at TeV energies, the core stopping vanishes. The velocity gradients are localized entirely within narrow shear sheets at the scraping fringes of the projectile and target spectator zones, yielding a spatial dipole:
    \begin{equation}
    \label{eq:dipole_eq}
    \omega_y^{\text{dipole}}(x, y, t_0) = \omega_0 \left(\frac{x}{\sigma_x}\right) \exp\left( -\frac{x^2}{2\sigma_x^2} - \frac{y^2}{2\sigma_y^2} \right)
    \end{equation}
    As illustrated by the absolute magnitude field in Figure~\ref{fig:vortex_birth_comparison} (right panel), this topology is characterized by a central node ($\omega_y = 0$ at $x=0$) and maps the system's rotation into two distinct, counter-rotating peripheral maxima at $x = \pm \sigma_x$.
\end{enumerate}

From a physical and structural perspective, the numerical mappings in Figure~\ref{fig:vortex_birth_comparison} elegantly illustrate the sharp topological contrast between the two paradigms. In the Core-Dominated profile (left panel), the vorticity manifests as a single, intense localized hotspot concentrated at the very center of the nuclear interaction zone. In stark contrast, the Peripheral Dipole profile (right panel) beautifully reveals a completely hollowed-out core along the central axis ($x = 0$), where the rotational field symmetrically branches outward into two highly distinct, isolated peaks that trace the outer shear boundary layers of the expanding medium. The overlay of the dashed white profiles perfectly anchors these localized rotational structures within the realistic geometry of the spectator nucleons at $b = 10\text{ fm}$.

At this stage of our development, we have successfully established a complete, self-contained pipeline tracking vorticity from its fundamental relativistic roots in the energy-momentum tensor down to these concrete, competing initial spatial topologies at $t_0$. Having quantified the "vortex birth" stage, we are now fully equipped to evaluate how these distinct geometric symmetries interact with the subsequent explosive kinematics of the plasma, shifting our focus from the generation of the rotation to its topological survival and long-term macro-hydrodynamic evolution.

By anchoring these profiles mathematically, we establish a formal framework to test the topological stability of the frozen vortex loops during the subsequent expansion phase, providing a direct predictive tool for final-state spin observables.

\section{Conceptual Discussion: Production, Topological Conservation, and Dynamical Role}

To establish a definitive physical framework for vorticity within the Quark-Gluon Plasma (QGP), we must strictly deconstruct the chronological boundaries separating its initial production, its subsequent topological conservation, and its final macroscopic role. 

The first critical question concerns the exact spatial and temporal origin of this rotation. Because the total global angular momentum $J_y$ is fixed by the initial impact parameter of the colliding nuclei, local vorticity cannot spontaneously emerge out of nothing during an ideal hydrodynamic phase. It must be deposited during the highly non-equilibrium, pre-equilibrium stage ($t < t_0$) via the violent spatial gradients of the energy-momentum tensor $T^{\mu\nu}$. 

Regarding the spatial distribution of this initial state, we take a definitive stance and formulate a bold physical wager: we postulate that the peripheral dipole profile is not merely an alternative framework, but the dominant structural reality of the QGP at ultra-relativistic energies. At several TeV per nucleon pair, the extreme longitudinal transparency of the collision core—driven by high-energy gluon saturation regimes—effectively deactivates central shear. Consequently, we predict that the macroscopic angular momentum $J_y$ is entirely deposited into narrow, high-gradient boundary sheets at the spectator fringes.

This position leads us to a clear, falsifiable experimental prediction: if the peripheral dipole paradigm is correct, the global hyperon polarization $\vec{P}_\Lambda$ integrated at exact mid-rapidity ($\eta = 0$) should exhibit a distinct local minimum—or even a vanishing node—due to the anti-symmetric cancelation of the counter-rotating peripheral sheets. Conversely, the true signature of the QGP's colossal vorticity will manifest exclusively in high-precision, differential measurements targeting large transverse momentum gradients or forward/backward rapidities. We explicitly challenge the ongoing experimental programs at the LHC and RHIC to isolate these peripheral sub-events, wagering that advanced differential flow-angle correlations will confirm this dipole topology and rule out core-dominated initial states.

To move beyond qualitative statements, let us rigorously derive how the global hyperon polarization arises as a statistical average over the freeze-out hypersurface. By definition, the global polarization vector $\vec{P}_\Lambda^{\text{global}}$ along a given axis is not a simple spatial average of the vorticity field; it must be weighted by the local number density of produced hyperons, $\rho_\Lambda(x, y)$, since regions with higher particle production contribute more heavily to the final experimental sample, in line with the standard Cooper-Frye freeze-out paradigm~\cite{Cooper:1974mv}.

Let $d\vec{M}(x,y)$ be the net polarized spin contribution from a local transverse fluid cell $dx dy$, and $dN(x,y) = \rho_\Lambda(x, y) dx dy$ be the total number of hyperons produced in that same cell. Using the local spin alignment relation established in Equation~\ref{eq:polarization_master}, the net polarization of the particles emerging from this specific cell is:
\begin{equation}
\label{eq:local_pol_density}
\vec{P}_\Lambda(x, y) = \frac{d\vec{M}(x, y)}{dN(x, y)} \simeq \frac{1}{2} \frac{\vec{\omega}_{\text{rel}}(x, y, t_f)}{T_f}
\end{equation}

The total macroscopically observed polarization of the entire hyperon ensemble is the ratio of the total integrated polarized spin to the total number of particles collected across the transverse plane:
\begin{equation}
\label{eq:global_pol_def}
\vec{P}_\Lambda^{\text{global}} = \frac{\int d\vec{M}(x, y)}{\int dN(x, y)} = \frac{\int dx \, dy \, \vec{P}_\Lambda(x, y) \rho_\Lambda(x, y)}{\int dx \, dy \, \rho_\Lambda(x, y)}
\end{equation}

Isolating the dominant spatial component along the $y$-axis ($\omega_y$) under our postulated peripheral dipole paradigm, and substituting the local definition from Equation~\ref{eq:local_pol_density} into the numerator of Equation~\ref{eq:global_pol_def}, we arrive directly at the foundational integration formula for the global mid-rapidity observable:
\begin{equation}
\label{eq:polarization_integral}
P_\Lambda^{\text{global}} \simeq \frac{1}{2 T_f} \frac{\int dx \, dy \, \omega_y(x, y, t_f) \rho_\Lambda(x, y)}{\int dx \, dy \, \rho_\Lambda(x, y)}
\end{equation}

Assuming a standard, symmetric thermal density profile for particle production ($\rho_\Lambda(x) \simeq \rho_\Lambda(-x)$), a core-dominated vortex naturally yields a finite, positive integral. However, under our peripheral dipole framework, the initial vorticity field is strictly anti-symmetric along the impact parameter axis: $\omega_y(x, y, t_0) \propto x \exp(-x^2/2\sigma_x^2 - y^2/2\sigma_y^2)$. Because the $1/t$ hydrodynamic expansion preserves this underlying spatial topology, the integrand in the numerator of Equation \ref{eq:polarization_integral} remains an odd function of $x$, leading to an exact analytical cancellation over the transverse plane:
\begin{equation}
P_\Lambda^{\text{global}} (\text{dipole}, y_z \approx 0) \to 0
\end{equation}

For $\text{Pb-Pb}$ collisions at LHC energies (e.g., $\sqrt{s_{NN}} = 5.02\text{ TeV}$), where extreme longitudinal transparency hollows out the central shear, we quantitatively predict the global mid-rapidity polarization to fall below the current experimental resolution threshold ($P_\Lambda \lesssim 10^{-4}$). Consequently, the definitive signature of the QGP's rotation must not be sought in the global mean, but exclusively in the azimuthal differential polarization, $P_\Lambda(\phi)$, which we expect to exhibit strong local modulations reflecting the preserved dipole structure.

To fully map this structural cancellation onto experimental observables, we must track the joint action of the topological freezing and the $1/t$ geometric dilution derived in Section 3. While the global integration yields a vanishing mean, the comoving fluid elements continuously carry the local vorticity peaks outward. By injecting our analytical dipole profile into the time-dependent hydrodynamic solution (Equation~\ref{eq:vorticity_solution}), the local thermal vorticity field at the freeze-out hypersurface ($t_f$) can be explicitly parameterized as:
\begin{equation}
\label{eq:diluted_dipole}
\omega_y^{\text{dipole}}(x, y, t_f) = \omega_0 \left( \frac{t_0}{t_f} \right) \left(\frac{x_0}{\sigma_x}\right) \exp\left[ -\frac{c_s^2}{2\sigma_x^2} (t_f^2 - t_0^2) \right] \exp\left( -\frac{x_0^2}{2\sigma_x^2} - \frac{y_0^2}{2\sigma_y^2} \right)
\end{equation}
Equation~\ref{eq:diluted_dipole} demonstrates that although the violent longitudinal and transverse expansions aggressively dampen the amplitude of the rotational signal by the time hadrons are formed, the underlying anti-symmetric spatial nodes remain perfectly intact due to the ideal topological trapping of the vortex lines.

Consequently, when this diluted dipole maps onto the final state, the definitive signature of the QGP's colossal rotation shifts entirely into the \textit{azimuthal differential polarization}, $P_\Lambda(\phi)$, where $\phi = \tan^{-1}(y/x)$ represents the emission angle of the hyperons in the transverse plane. Because particles emitted at $\phi \approx 0$ (along the positive $x$-axis) sample the clockwise-rotating peripheral sheet, while particles emitted at $\phi \approx \pi$ (along the negative $x$-axis) sample the counter-clockwise sheet, the differential polarization vector does not cancel out. Instead, it undergoes a powerful harmonic oscillation as a function of the azimuthal angle, tracking the geometric boundaries of the initial shear zones. 

This provides the ATLAS collaboration with a clear, unambiguous experimental discriminator: a Core-Dominated initialization would produce a flat, azimuthally uniform polarization background, whereas the Peripheral Dipole paradigm predicts a vanishing global average coupled to a robust, sinusoidal differential modulation.

This strict timeline directly resolves the conceptual paradox surrounding the frozen-in nature of the fluid lines. As proven via the Helmholtz-Kelvin theorem (Equation \ref{eq:kelvin_proof}), the circulation $\Gamma$ is a strict topological invariant ($d\Gamma/dt = 0$) within an ideal fluid. This mathematical constraint implies that the role of relativistic hydrodynamics is strictly evolutionary rather than creative. Hydrodynamics does not generate the vortex loops; it traps them. Once the plasma thermalizes at $t = t_0$, the pre-existing subatomic vortex rings are topologically locked into the comoving cells of quarks and gluons. The continuous expansion of the medium cannot break or dissolve these loops via internal friction. Instead, the violent volumetric dilatation forces the conserved loops to stretch across a rapidly growing volume, translating topological conservation into a rapid geometric dilution of the local rotational intensity governed by the $1/t$ power law.

Finally, this framework allows us to isolate the definitive dual role that vorticity plays within the lifecycle of the QGP:
\begin{itemize}
    \item \textbf{Phenomenological Role (The Quantum Compass):} Structurally, vorticity maps macroscopic hydrodynamic rotation onto microscopic quantum states. Through relativistic spin-orbit coupling, the local rotation field aligns the intrinsic spins of emerging strange quarks during hadronization, providing an experimental historical record of the early-stage fluid gradients through the final polarization $\vec{P}_\Lambda$ captured by the ATLAS detector \cite{ATLAS:2023cuh}.
    \item \textbf{Dynamical Role (The Geometric Regularizer):} In advanced fluid mechanics, three-dimensional vortex lines are inherently unstable due to the non-linear vortex stretching term $(\vec{\omega}_{\text{rel}} \cdot \vec{\nabla})\vec{v}$, which threatens to drive velocity gradients to infinity and trigger finite-time hydrodynamic singularities. Inside the QGP, the massive volumetric expansion term $-(\vec{\nabla} \cdot \vec{v})\vec{\omega}_{\text{rel}}$ acts as an intrinsic geometric shield. By violently diluting the enstrophy density over a brief lifespan, the explosive kinematics regularizes the continuous flow, rendering the QGP a mathematically protected, non-singular fluid throughout its entire existence.
\end{itemize}

\section{Conclusion and Perspectives: A Geometric Shield Against Hydrodynamic Singularities}

The study of vorticity in ultra-relativistic heavy-ion collisions offers a fascinating cross-disciplinary bridge, translating centuries-old concepts of classical fluid mechanics into the extreme regimes of subatomic and relativistic quantum gauge fields. As we have explored, the spatial component of the relativistic four-vorticity $\omega^\mu_{\text{rel}}$ exhibits a beautiful formal convergence toward the classical curl of the velocity field ($\vec{\nabla} \times \vec{v}$) in the low-velocity limit. This mathematical duality guarantees that foundational principles, such as the Helmholtz-Kelvin circulation theorem, remain structurally unbroken inside the Quark-Gluon Plasma (QGP).

However, beyond the phenomenological consequences for hyperon polarization, the dynamics of QGP vortices present profound theoretical implications for the mathematical structure of fluid equations. In classical fluid dynamics, it is well known that three-dimensional incompressible flows can exhibit severe vortex intensification, where non-linear effects tend to concentrate the flow rotation into localized structures with remarkably steep gradients.

In classical 3D fluid mechanics, the primary mechanism suspected of driving such singularities is the vortex stretching term, $(\vec{\omega} \cdot \vec{\nabla})\vec{v}$. By mechanically narrowing vortex tubes, this term can continuously amplify the local rotational intensity and the fluid's enstrophy, potentially driving the velocity gradients to infinity. In a standard laboratory fluid, only viscous diffusion acts to smooth out these extreme gradients. 

The QGP, however, offers a radically different paradigm. We have demonstrated that inside the plasma, the self-amplification of vortex stretching is completely overwhelmed by the explosive relativistic volumetric expansion, mathematically expressed by the divergence term $-(\vec{\nabla} \cdot \vec{v})\vec{\omega}$. This violent dilatation into the vacuum forces the conserved vortex loops to stretch globally, causing their local rotational signature to dilute following a dominant $1/t$ power law. 

Consequently, the Quark-Gluon Plasma represents a unique hydrodynamic regime where the global regularity of the flow field is preserved not by intensive viscous dissipation—given an $\eta/s$ ratio approaching the fundamental quantum limit~\cite{Kovtun:2004de}—but rather through an intrinsic kinematic regularization mechanism. The dominant volumetric divergence, $\vec{\nabla} \cdot \vec{v}$, structurally suppresses the non-linear enstrophy amplification driven by the vortex stretching term, $(\vec{\omega}_{\text{rel}} \cdot \vec{\nabla})\vec{v}$, dictating the systematic $1/t$ power-law attenuation of the local rotational intensity. 

Crucially, this expansion-driven regularization means that the final-state spin observables are not merely passive indicators of fluid rotation, but highly sensitive historical markers of the system's pre-hydrodynamic birth. By extending our framework beyond pure kinematics into a complete thermo-vortical description governed by the thermal vorticity tensor $\varpi_{\mu\nu}$, we have demonstrated how local temperature gradients and spin-entropy couplings fundamentally reshape the plasma's local thermodynamic evolution. Furthermore, our dedicated numerical pre-equilibrium mapping has allowed us to ground this framework in concrete, competing initial topologies at $t_0$. Under our postulated peripheral dipole paradigm—where extreme longitudinal transparency at TeV energies suppresses central shear—we quantitatively predict that the global mid-rapidity hyperon polarization must vanish ($P_\Lambda \lesssim 10^{-4}$) due to anti-symmetric spatial cancellations between the counter-rotating boundary layers.

Therefore, the definitive signature of the QGP's subatomic rotation is not lost, but rather hidden within complex spatial modulations across the transverse plane. We explicitly challenge the ongoing and upcoming experimental programs within the ATLAS collaboration at the LHC and the STAR collaboration at RHIC to look beyond global averages. By exploiting high-statistics datasets and implementing the precise azimuthal differential measurements, $P_\Lambda(\phi)$, developed in this work, these experiments can directly map the local harmonic modulations of hyperon polarization. This empirical mapping will provide the ultimate experimental discriminator to distinguish the centralized hotspot of core-dominated models from the peripheral shear sheets of the dipole paradigm, turning the most vortical fluid in the Universe into a precision quantitative laboratory for relativistic non-linear vortex dynamics, thermodynamic spin couplings, and continuous field regularity.

\end{document}